\documentclass[%
a4paper,%
12pt,%
]{article}

\usepackage[dvips]{graphicx}

\title{The method of multiple internal reflections in a description
of tunneling evolution of nonrelativistic particles and photons}

\author{
Sergei~P.~Maydanyuk\thanks{E-mail: maidan@kinr.kiev.ua},
Vladislav~S.~Olkhovsky\thanks{E-mail: olkhovsk@kinr.kiev.ua}\\
and
Alexander~K.~Zaichenko\thanks{E-mail: zaichenk@kinr.kiev.ua}\\
\small\emph{Institute for Nuclear Research,
National Academy of Sciences of Ukraine,} \\
\small\emph{prosp. Nauki, 47, Kiev-28, 03680, Ukraine}}

\date{}

\begin{document}
\begin{sloppypar}
\maketitle

\begin{abstract}
A non-stationary method for tunneling description of
non-relativistic particles and photons through a barrier on the
basis of consideration of the multiple internal reflections of
vawe packets in relation of barrier boundaries is presented.
The method is described in details and proved in the case of the
one-dimentional tunneling of the particle through the rectangular
barrier.
For problems of the tunneling of the particle through the spherically
symmetric barrier and of the photon through the one-dimensional
barrier the amplitudes of transmitted and reflected wave packets
in relation to the barrier, times of the tunneling and the reflection
are found using of the method.
Hartman's and Fletcher's effect is analysed.
\end{abstract}

{\bf PACS number(s): 03.40.Kf, 03.50.De, 03.65.Nk, 41.20.Jb}

{\bf Keywords:}
1D- and 3D-tunneling,
spherically symmetric elastic scattering,
multiple internal reflections,
wave packet,
dynamics,
transmission and reflection coefficients,
tunneling and reflection times,
Hartman's and Fletcher's effect

{\bf UDC 539.14}

\section{Introduction
\label{sec:level1}}

The approach for description of a propagation of a nonrelativistic
particle above a barrier using the account of multiple internal
reflections of stationary plane waves between barrier boundaries
which describe a motion of this particle in the region of the
barrier, was considered in series of articles and is known for
a long time \cite{Fermor.1966.AJPIA,McVoy.1967.RMPHA,Anderson.1989.AJPIA}.
Thus, stationary solutions were studied only in the previous articles.

To apply this approach for a solution of a problem, one can need in
expressions for a wave function (w.f.) in the region of the barrier
to separate components having fluxes, directed in opposite
sides. For a problem of the particle propagating above the barrier it
appears primely enough. So, considering an one-dimensional (1D)
rectangular barrier, plane waves $e^{\pm ikx}$ can be taken as
such solutions, where $k$ is a wave vector.
To receive solutions using this approach for a problem of
the tunneling of the particle under such barrier
it appears more complicatedly, because in a consideration of
the tunneling as a stationary process the decreasing and increasing
components of the stationary w.f. in dependence on $x$ (being
analytic continuations of relevant expressions of the waves for the
case of above-barrier energies) in a sub-barrier region correspond
to zero fluxes separately and it is not correct to use them as
the propagating waves from a physical point of view.
If to define expressions for such waves by another way (for example,
having required an existence of the nonzero fluxes in the barrier region
at each step of this approach), then we receive a divergence of
the expressions for the waves for the above-barrier and the sub-barrier cases.
However, the flux calculated on a basis of a complete stationary
w.~f., is not equal to zero, and, therefore, the tunneling of the
particle under the barrier exists.

As a further development of a time analysis of tunneling processes
submitted in articles
\cite{Olkhovsky.1992.PRPLC,Olkhovsky.1997.Trieste,Olkhovsky.1997.JPGCE},
here we represent the non-stationary solution method of a problem of
tunneling of a nonrelativistic particle or a photon through a barrier
using multiple internal reflections of fluxes in a sub-barrier region
in relation to barrier boundaries (we name this approach as the
method of multiple internal reflections). In given article we study
an one-dimensional and a spherically symmetric problems.

At analysing the tunneling (or the propagation) of the nonrelativistic
particle, an important specific feature of this method is a description
of a particle motion using non-stationary wave packets (w.p.).
Due to this one can determine correctly the packets
propagating in different directions in a barrier region, fulfil
a time analysis of the particle tunneling (propagation) and study in
details this process in an interesting time moment or in relation to
a concrete point of space. For obtaining time parameters of
the tunneling, this method has shown itself convenient and simply enough.

For the problem with a spherically symmetric barrier the reflected and
transmitted w.p. in relation to this barrier are propagate in one
direction. Stationary solution methods do not allow to separate
the w.p., transmitted through the barrier, from the w.p., reflected
from the barrier. Using the method of multiple internal reflections,
one can find amplitudes and expressions for these w.p..
In result, it appears possible to present S-matrix in a form of
a sum of two components corresponding to stationary parts for the reflected
and transmitted w.p. in relation to the barrier. The sum of the stationary
parts for these w.p., obtained by this method,
converges with the expression for a scattered wave obtained by an usual
stationary method. The spherically symmetric problem with use of given
approarch is considered for the first time.

At first we consider the problem of the tunneling of the
nonrelativistic particle through the one-dimensional rectangular
barrier.
This problem is a test one and allows to analyse specific features
of this method.

Further the problem of the tunneling of the particle through
the spherically symmetric barrier, which radial part has a rectangular
form, is solved. For it amplitudes of the transmitted and reflected
w.p., total times of the tunneling and reflection in relation to the
barrier are found. Hartman's and Fletcher's effect is analysed.
An expression for S-matrix is presented in a form of a sum of two
components corresponding to amplitudes of stationary parts for
the transmitted and reflected w.p..
The time parameters using of the method of multiple internal
reflections are found for the first time.

One can apply the method to a problem solution of the tunneling of the
particle through a spherically symmetric barrier, which radial part
has an arbitrary form, if a general stationary solution for a w.f.
is known for this potential. Some problems with various barrier forms
are considered. The problems are selected so that to show better 
features of the method at their solution.

At a finishing of the article a possibility to use the method in a
problem of a tunneling of photons through an one-dimensional
rectangular barrier is studied.
On a basis of a given analysis the method is proved for
the problem with the photons. Using
a found transformation the results and solutions of the problem
with the nonrelativistic particle transform into corresponding
expressions for the problem with the photons. Hartman's and Fletcher's
effect is analysed.

\section{Tunneling of a particle through an one-dimensional
rectangular barrier
\label{sec:level2}}

\begin{figure}[p]
\centerline{\includegraphics[width=7cm]{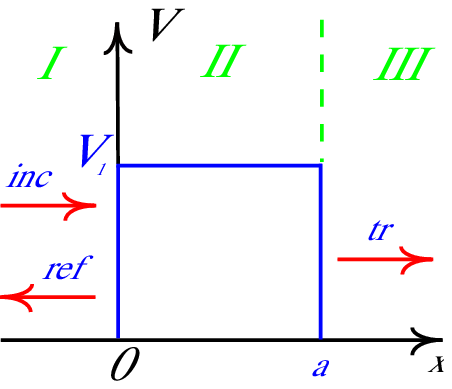}}
\caption{One-dimensional rectangular barrier}
\label{fig1}
\end{figure}

Let's consider a problem of tunneling of a nonrelativistic particle
in a positive $x$-dirrection through an one-dimensional rectangular
potential barrier (see Fig.\ \ref{fig1}).
Let's label a region I for $x < 0$, a region II for
$0 < x < a$ and a region III for $x > a$, accordingly.
Let's study an evolution of its tunneling through the barrier.


In a beginning we consider a standard approach to a solution of this
problem \cite{Landau.1974,Babicov.1988}. Let's consider a case when 
levels of energy lay under than a height of the barrier: $E < V_{1}$.

The tunneling evolution of the particle can be described using
a non-stationary consideration of a propagating w.p.
\begin{equation}
  \psi(x, t) = \int\limits_{0}^{+\infty} g(E - \bar{E}) \varphi(k, x)
                e^{-iEt/\hbar} dE,
\label{eq2_1}
\end{equation}                                  
where the stationary w.f. has a form:
\begin{equation}
\varphi(x) = \left\{
\begin{array}{ll}
   e^{ikx}+A_{R}e^{-ikx},               & \mbox{for } x<0;   \\
   \alpha e^{\xi x} + \beta e^{-\xi x}, & \mbox{for } 0<x<a; \\
   A_{T} e^{ikx},                       & \mbox{for } x>a;
\end{array} \right.
\label{eq2_2}
\end{equation}                                  
and $k   = \frac{1}{\hbar}\sqrt{2mE}$,
    $\xi = \frac{1}{\hbar}\sqrt{2m(V_{1}-E)}$,
$E$ and $m$ are the total energy and mass of the particle, accordingly.
The weight amplitude $g(E - \bar{E})$ can be written in a form of
gaussian \cite{Olkhovsky.1992.PRPLC} and satisfies to a requirement
of the normalization $\int |g(E - \bar{E})|^{2} dE = 1$, value $\bar{E}$
is an average energy of the particle.
One can calculate coefficients $A_{T}$, $A_{R}$, $\alpha$ and
$\beta$ analytically, using a requirements of a continuity of w.f.
$\varphi(x)$ and its derivative on each boundary of the barrier.

Substituting in Eq.\ (\ref{eq2_1}) instead of $\varphi(k, x)$ the
incident $\varphi_{inc}(k, x)$, transmitted  $\varphi_{tr}(k, x)$
or reflected part of w.f. $\varphi_{ref}(k, x)$, defined by
Eq.\ (\ref{eq2_2}), we receive the incident, transmitted  or reflected
w.p., accordingly.


We assume, that a time, for which the w.p. tunnels through the barrier,
is enough small. So, the time necessary for a tunneling of an
$\alpha$-partiále through a barrier of decay in $\alpha$-decay
of a nucleus, is about $10^{-21}$ sekonds \cite{Kasagi.1997.PRLTA}.
We consider, that one can neglect a spreading of the w.p. for
this time. And a breadth of the w.p. appears essentially more narrow
on a comparison with a barrier breadth
\cite{Olkhovsky.1992.PRPLC,Olkhovsky.1997.Trieste,Olkhovsky.1997.JPGCE}.
Considering only sub-barrier processes, we exclude a component of waves
for above-barrier energies, having included the additional transformation
\begin{equation}
   g(E - \bar{E}) \to g(E - \bar{E}) \theta(V_{1} - E),
\label{eq2_3}
\end{equation}                                  
where $\theta$-function satisfies to the requirement
%
\[
\theta(\eta) = \left\{
\begin{array}{ll}
   0,               & \mbox{for } \eta<0;   \\
   1,               & \mbox{for } \eta>0.
\end{array} \right.
\]

The method of multiple internal reflections considers the
propagation process of the w.p. describing a motion of the particle,
sequentially on steps of its penetration in relation to each
boundary of the barrier
\cite{Fermor.1966.AJPIA,McVoy.1967.RMPHA,Anderson.1989.AJPIA}. Using
this method, we find expressions for the transmitted and
reflected w.p. in relation to the barrier.

At the first step we consider the w.p. in the region I, which is
incident upon the first (initial) boundary of the barrier. Let's
assume, that this package transforms into the w.p., transmitted
through this boundary and tunneling further in the region II, and into
the w.p., reflected from the boundary and propagating back in the region
I. Thus we consider, that the w.p., tunneling in the region II, is not
reached the second (final) boundary of the barrier because of a
terminating velocity of its propagation, and consequently at this
step we consider only two regions I and II. Because of physical
reasons to construct an expression for this packet, we consider,
that its amplitude should decrease in a positive $x$-direction. We
use only one item $\beta\exp(-\xi x)$ in Eq.\ (\ref{eq2_2}), throwing
the second increasing item $\alpha\exp(\xi x)$ (in an opposite
case we break a requirement of a finiteness of the w.f. for an
indefinitely wide barrier). In result, in the region II we obtain:
\begin{equation}
  \psi^{1}_{tr}(x, t) = \int\limits_{0}^{+\infty} g(E - \bar{E})
  \theta(V_{1} - E) \beta^{0} e^{-\xi x -iEt/\hbar} dE,
  \mbox{for } 0<x<a.
\label{eq2_4}
\end{equation}                                  
Thus the w.f. in the barrier region constructed by such way, is an
analytic continuation of a relevant expression for the w.f.,
corresponding to a similar problem with above-barrier
energies, where as a stationary expression we select the wave
$\exp(ik_{2}x)$, propagated to the right.

Let's consider the first step further. One can write expressions
for the incident and the reflected w.p. in relation to the first boundary
as follows
\begin{equation}
\begin{array}{lcll}
\psi_{inc}(x, t) & = & \int\limits_{0}^{+\infty} g(E - \bar{E})
        \theta(V_{1} - E) e^{ikx -iEt/\hbar} dE,
        & \mbox{for } x<0, \\
\psi^{1}_{ref}(x, t) & = & \int\limits_{0}^{+\infty} g(E - \bar{E})
        \theta(V_{1} - E) A_{R}^{0} e^{-ikx -iEt/\hbar} dE,
        & \mbox{for } x<0.
\end{array}
\label{eq2_5}
\end{equation}                                  
A sum of these expressions represents the complete w.f. in the region
I, which is dependent on a time. Let's require, that this w.f. and its
derivative continuously transform into the w.f. (\ref{eq2_4}) and its
derivative at point $x=0$
(we assume, that the weight amplitude $g(E - \bar{E})$ differs weakly
at transmitting and reflecting of the w.p. in relation to the barrier
boundaries).
In result, we obtain two equations, in
which one can pass from the time-dependent w.p. to the corresponding
stationary w.f. and obtain the unknown coefficients $\beta^{0}$
and $A_{R}^{0}$.

At the second step we consider the w.p., tunneling in the region II and
incident upon the second boundary of the barrier at point $x = a$.
It transforms into the w.p., transmitted through this boundary and
propagated in the region III, and into the w.p., reflected from the boundary
and tunneled back in the region II.
For a determination of these packets one can use Eq.\ (\ref{eq2_1}) with
account (\ref{eq2_3}), where as the stationary w.f. we use:
\begin{equation}
\begin{array}{lcll}
\varphi_{inc}^{2}(k, x) & = & \varphi_{tr}^{1}(k, x) =
        \beta^{0} e^{-\xi x},
        & \mbox{for } 0<x<a, \\
\varphi_{tr}^{2}(k, x) & = & A_{T}^{0}e^{ikx},
        & \mbox{for } x>a, \\
\varphi_{ref}^{2}(k, x) & = & \alpha^{0} e^{\xi x},
        & \mbox{for } 0<x<a.
\end{array}
\label{eq2_6}
\end{equation}                                  
Here, for forming an expression for the w.p. reflected from the
boundary, we select an increasing part of the stationary solution 
$\alpha^{0} \exp(\xi x)$ only. Imposing a condition of continuity
on the time-dependent w.f. and its derivative at point $x = a$, we
obtain 2 new equations, from which we find the unknowns coefficients
$A_{T}^{0}$ and $\alpha^{0}$.

At the third step the w.p., tunneling in the region II, is incident upon
the first boundary of the barrier. Then it transforms into the w.p.,
transmitted through this boundary and propagated further in the region
I, and into the w.p., reflected from boundary and tunneled back in
the region II. For a determination of these packets one can use Eq.\
(\ref{eq2_1}) with account Eq.\ (\ref{eq2_3}), where as the stationary w.f.
we use:
\begin{equation}
\begin{array}{lcll}
\varphi_{inc}^{3}(k, x) & = & \varphi_{ref}^{2}(k, x),
        & \mbox{for } 0<x<a, \\
\varphi_{tr}^{3}(k, x) & = & A_{R}^{1}e^{-ikx},
        & \mbox{for } x<0, \\
\varphi_{ref}^{3}(k, x) & = & \beta^{1} e^{-\xi x},
        & \mbox{for } 0<x<a.
\end{array}
\label{eq2_7}
\end{equation}                                  
Using a conditions of continuity for the time-dependent w.f. and
its derivative at point $x = 0$, we obtain the unknowns coefficients
$A_{R}^{1}$ and $\beta^{1}$.

Analysing further possible processes of the transmission (and the reflection)
of the w.p. through the boundaries of the barrier, we come to a deduction,
that any of following steps can be reduced to one of 2 considered
above. For the unknown coefficients $\alpha^{n}$, $\beta^{n}$,$A_{T}^{n}$
and $A_{R}^{n}$, used in expressions for the w.p., forming in result of
some internal reflections from the boundaries, one can obtain the
recurrence relations:
\begin{equation}
\begin{array}{lll}
\beta^{0} = \displaystyle\frac{2k}{k+i\xi},     &
\alpha^{n} = \beta^{n} \displaystyle\frac{i\xi-k}{i\xi+k}e^{-2\xi a}, &
\beta^{n+1} = \alpha^{n} \displaystyle\frac{i\xi-k}{i\xi+k}, \\
A_{R}^{0} = \displaystyle\frac{k-i\xi}{k+i\xi},     &
A_{T}^{n} = \beta^{n} \displaystyle\frac{2i\xi}{i\xi+k}e^{-\xi a-ika}, &
A_{R}^{n+1} = \alpha^{n} \displaystyle\frac{2i\xi}{i\xi+k}.
\end{array}
\label{eq2_8}
\end{equation}                                  

Considering the propagation of the w.p. by such way, we obtain 
expressions for the w.f. on each region which can be written through
series of multiple w.p.. Using Eq.\ (\ref{eq2_1}) with account
Eq.\ (\ref{eq2_3}), we determine resultant expressions for the
incident, transmitted and reflected w.p. in relation to the barrier,
where one can need to use following expressions for the stationary
w.f.:
\begin{equation}
\begin{array}{lcll}
\varphi_{inc}(k, x) & = & e^{ikx},
                        & \mbox{for } x<0, \\
\varphi_{tr}(k, x)  & = & \sum\limits_{n=0}^{+\infty} A_{T}^{n} e^{ikx}, 
                        & \mbox{for } x>a, \\
\varphi_{ref}(k, x) & = & \sum\limits_{n=0}^{+\infty} A_{R}^{n} e^{-ikx}, 
                        & \mbox{for } x<0.
\end{array}
\label{eq2_9}
\end{equation}                                  

Now we consider the w.p. formed in result of sequential $n$ reflections
from the boundaries of the barrier and incident upon one of these
boundaries at point $x = 0$ ($i = 1$) or at point $x = a$ ($i = 2$).
In result, this w.p. transforms into the w.p. $\psi_{tr}^{i}(x, t)$,
transmitted through boundary with number $i$, and into the w.p.
$\psi_{ref}^{i}(x, t)$, reflected from this boundary. For an
independent on $x$ parts of the stationary w.f. one can write:
\begin{equation}
\begin{array}{ll}
   \displaystyle\frac{\varphi_{tr}^{1}}{\exp(-\xi x)} =
   T_{1}^{+} \displaystyle\frac{\varphi_{inc}^{1}}{\exp(ikx)}, &
   \displaystyle\frac{\varphi_{ref}^{1}}{\exp(-ikx)} =
   R_{1}^{+} \displaystyle\frac{\varphi_{inc}^{1}}{\exp(ikx)}, \\
   \displaystyle\frac{\varphi_{tr}^{2}}{\exp(ikx)} =
   T_{2}^{+} \displaystyle\frac{\varphi_{inc}^{2}}{\exp(-\xi x)}, &
   \displaystyle\frac{\varphi_{ref}^{2}}{\exp(\xi x)} =
   R_{2}^{+} \displaystyle\frac{\varphi_{inc}^{2}}{\exp(-\xi x)}, \\
   \displaystyle\frac{\varphi_{tr}^{1}}{\exp(-ikx)} =
   T_{1}^{-} \displaystyle\frac{\varphi_{inc}^{1}}{\exp(\xi x)}, &
   \displaystyle\frac{\varphi_{ref}^{1}}{\exp(-\xi x)} =
   R_{1}^{-} \displaystyle\frac{\varphi_{inc}^{1}}{\exp(\xi x)}, 
\end{array}
\label{eq2_10}
\end{equation}                                  
where the sign "+" (or "-") corresponds to the w.p., tunneling (or
propagating) in a positive (or negative) $x$-direction and incident
upon the boundary with number $i$. Using $T_{i}^{\pm}$ and $R_{i}^{\pm}$,
one can precisely describe an arbitrary w.p. which has formed in result
of $n$-multiple reflections, if to know a ``path'' of its propagation
along the barrier. Using the recurrence relations Eq.\ (\ref{eq2_8}),
the coefficients $T_{i}^{\pm}$ and $R_{i}^{\pm}$ can be obtained.
\begin{equation}
\begin{array}{lll}
T_{1}^{+} = \beta^{0},                           
&
T_{2}^{+} = \displaystyle\frac{A_{T}^{n}}{\beta^{n}},
&
T_{1}^{-} = \displaystyle\frac{A_{R}^{n+1}}{\alpha^{n}},
\\
R_{1}^{+} = A_{R}^{0},
&
R_{2}^{+} = \displaystyle\frac{\alpha^{n}}{\beta^{n}},
&
R_{1}^{-} = \displaystyle\frac{\beta^{n+1}}{\alpha^{n}}.
\end{array}
\label{eq2_11}
\end{equation}                                  

Using the recurrence relations, one can find series of
coefficients $\alpha^{n}$, $\beta^{n}$, $A_{T}^{n}$ and $A_{R}^{n}$.
However, these series can be calculated easier, using coefficients
$T_{i}^{\pm}$ and $R_{i}^{\pm}$. Analysing all possible ``paths''
of the w.p. propagations along the barrier, we receive:
\begin{equation}
\begin{array}{lcl}
\sum\limits_{n=0}^{+\infty} A_{T}^{n} & = &
        T_{2}^{+}T_{1}^{-} \biggl(1 + \sum\limits_{n=1}^{+\infty}
                                 (R_{2}^{+}R_{1}^{-})^{n} \biggr) =
        \displaystyle\frac{i4k \xi e^{-\xi a-ika}}{F_{sub}},  \\
\sum\limits_{n=0}^{+\infty} A_{R}^{n} & = &
        R_{1}^{+} + T_{1}^{+}R_{2}^{+}T_{1}^{-} \biggl(1 +
        \sum\limits_{n=1}^{+\infty}(R_{2}^{+}R_{1}^{-})^{n} \biggr) =
        \displaystyle\frac{k_{0}^{2}D_{-}}{F_{sub}},         \\
\sum\limits_{n=0}^{+\infty} \alpha^{n} & = &
        \alpha^{0} \biggl(1 + \sum\limits_{n=1}^{+\infty}
                           (R_{2}^{+}R_{1}^{-})^{n} \biggr) =
        \displaystyle\frac{2k(i\xi - k)e^{-2\xi a}}{F_{sub}}, \\
\sum\limits_{n=0}^{+\infty} \beta^{n} & = &
        \beta^{0} \biggl(1 + \sum\limits_{i=1}^{+\infty}
                          (R_{2}^{+}R_{1}^{-})^{n} \biggr) =
        \displaystyle\frac{2k(i\xi + k)}{F_{sub}}, \\
\end{array}
\label{eq2_12}
\end{equation}                                  
where
\begin{equation}
\begin{array}{lll}
F_{sub} & = & (k^{2} - \xi^{2})D_{-} + 2ik\xi D_{+},            \\
D_{\pm}   & = & 1 \pm e^{-2\xi a},                              \\
k_{0}^{2} & = & k^{2} + \xi^{2} = \displaystyle\frac{2mV_{1}}{\hbar^{2}}.
\end{array}
\label{eq2_13}
\end{equation}                                  

All series $\sum \alpha^{n}$, $\sum \beta^{n}$, $\sum A_{T}^{n}$
and $\sum A_{R}^{n}$, obtained using the method of multiple internal
reflections, coincide with the corresponding coefficients $\alpha$,
$\beta$, $A_{T}$ and $A_{R}$ of the Eq.\ (\ref{eq2_2}), calculated
by a stationary methods
\cite{Olkhovsky.1992.PRPLC,Landau.1974,Razavy.1988.PRPLC}.
Using the following substitution
\begin{equation}
i\xi \to k_{2},
\label{eq2_14}
\end{equation}                                  
where $k_{2}= \frac{1}{\hbar}\sqrt{2m (E-V_{1})}$ is a wave number
for a case of above-barrier energies, expression for the coefficients
$\alpha^{n}$, $\beta^{n}$, $A_{T}^{n}$ and $A_{R}^{n}$ for each step,
expressions for the w.f. for each step, the total Eqs.\ (\ref{eq2_12}) and
(\ref{eq2_13}) transform into the corresponding expressions for a
problem of the particle propagation above this barrier. At the
transformation of the w.p. and the time-dependent w.f. one can need to
change a sign of argument at $\theta$-function. Besides the following
property is fulfilled:
\begin{equation}
\biggl|\sum\limits_{n=0}^{+\infty} A_{T}^{n}\biggr|^{2} +
\biggl|\sum\limits_{n=0}^{+\infty} A_{R}^{n}\biggr|^{2} = 1.
\label{eq2_15}
\end{equation}                                  

\section{Tunneling of the particle through a spherically symmetric
rectangular barrier
\label{sec:level4}}

\subsection{Transmitted and reflected wave packets}

A problem of a motion of two interacting particles can be
redused to a problem of one particle scattering in a spherically
symmetric field.
Let's assume, that the particle under an action of a central force
\begin{equation}
V(r) = \left\{
\begin{array}{rll}
  -V_{0}, & \mbox{for } r<R_{1};       & \mbox{(region I)};  \\
   V_{1}, & \mbox{for } R_{1}<r<R_{2}, & \mbox{(region II)}; \\
   0,     & \mbox{for } r>R_{2},       & \mbox{(region III)}.
\end{array} \right.
\label{eq4_1}
\end{equation}                                  
is incident outside upon an external boundary of the barrier at point
$r=R_{2}$ (see Fig. 2).

Let's study an evolution of tunneling of the particle through the barrier.
We consider a case when the moment $l = 0$ and levels of energy
lay below than a barrier height.
The tunneling evolution of the particle in time
dependence can be described using a w.p. constructed on a
basis of a stationary solution of the following form \cite{Landau.1974}:
\begin{equation}
\psi(r, \theta, \varphi) =  \frac{\chi(r)}{r}
   Y_{lm}(\theta, \varphi),
\label{eq4_2}
\end{equation}                                  
\begin{equation}
\chi(r) = \left\{
\begin{array}{lll}
A(e^{-ik_{1}r} - e^{ik_{1}r}), & \mbox{for }  r<R_{1},
        & \mbox{(region I)}, \\
\alpha e^{\xi r} + \beta e^{-\xi r}, & \mbox{for } R_{1}<r<R_{2},
        & \mbox{(region II)}, \\
e^{-ikr} + Se^{ikr}, & \mbox{for } r>R_{2}, & \mbox{(region III)},
\end{array} \right.
\label{eq4_3}
\end{equation}                                  
where $Y_{lm}(\theta, \varphi)$ is a spherical function,
    $k_{1} = \frac{1}{\hbar}\sqrt{2m(E+V_{0})}$, 
    $\xi = \frac{1}{\hbar}\sqrt{2m(V_{1}-E)}$,
    $k = \frac{1}{\hbar}\sqrt{2mE}$.
For the spherically symmetric problem in a case of sub-barrier
energies we obtain:
%
\begin{equation}
\chi(r, t) = \int\limits_{0}^{+\infty} g(E - \bar{E})
             \theta(V_{l} - E) \chi(k, r) e^{-iEt/\hbar} dE,
\label{eq4_4}
\end{equation}                                  
\begin{equation}
V_{l}(r)   = V(r) + \displaystyle\frac{\hbar^{2}}{2m}
             \displaystyle\frac{l(l+1)}{r^{2}},
\label{eq4_5}
\end{equation}                                  
%
where the second item in Eq.\ (\ref{eq4_5}) is a centrifugal
energy, which is equal to zero at $l = 0$, weight amplitude
$g(E - \bar{E})$ and average energy of the particle $\bar{E}$
are defined similarly to the one-dimensional problem (see Sec.
\ref{sec:level2}).

At a stationary consideration of the solutions (\ref{eq4_3}) we
describe the particle incident upon the external boundary of
the barrier by a spherical wave $\exp(-ikr)$ convergent to the centre.
And we describe the particle scattered on the barrier in the region
III by a spherical wave $S \exp(ikr)$ divergent outside.
The scattered wave takes into account both a possibility of
a reflection of the particle from the barrier, which is written by the
divergent wave, and a possibility of a penetration of the particle through
the barrier, when in a beginning the particle tunnels from the region III
to the region I, and then after some period of time it tunnels back from
the region I to the region III and also is written by the divergent wave.
Only one item $S \exp(ikr)$ contains the transmitted and reflected
divergent waves, and it is impossible to separate them at the stationary
consideration.

As non-stationary, the method of multiple internal reflections allows
to find a solution of this problem. Let's apply it to this problem.
We study a propagation of a w.p. describing tunneling of the particle,
sequentially on steps of its transmission in relation to each of
boundaries of the barrier (similarly to the one-dimensional problem).
In result of an analysis we come to a deduction, that any step in
such viewing of the propagation of the w.p. along the barrier will be
similar to one of 4 steps independent among themselves. Analysing
these 4 steps further, one can obtain recurrence relations for
finding coefficients $A^{n}$, $S^{n}$, $\alpha^{n}$ and
$\beta^{n}$ for an arbitrary step $n$.

In result of multiple internal reflections (and transitions) in
relation to the boundaries of the barrier a total time-dependent
w.f. in each region can be written in a form of series, composed
from convergent and divergent w.p.. Analysing possible ``paths''
of propagations of these packets, one can calculate expressions
for series of coefficients $S^{n}$, $A^{n}$, $\alpha^{n}$ and
$\beta^{n}$:
\begin{equation}
\begin{array}{lcl}
\sum\limits_{n=1}^{+\infty} S^{n} & = &
     \displaystyle\frac{1}{F_{sub}}
     T_{2}^{-}T_{2}^{+} (R_{1}^{-}(1-R_{1}^{+}R_{0}^{-}) +
     T_{1}^{-}R_{0}^{-}T_{1}^{+}) = \\
 & = & \displaystyle\frac{4ik\xi
     \biggl(\displaystyle\frac{i\xi-k_{1}}{i\xi+k_{1}} -
     e^{2ik_{1}R_{1}}\biggr) e^{2\xi(R_{1}-R_{2}) - 2ikR_{2}} }
     {F_{sub} (k+i\xi)^{2}}, \\
\sum\limits_{n=0}^{+\infty}A^{n} & = &
    \displaystyle\frac{T_{1}^{-}T_{2}^{-}}{F_{sub}} =
    \displaystyle\frac{4ik\xi e^{-ikR_{2} + ik_{1}R_{1} -
    \xi(R_{2}-R_{1})}} {F_{sub} (k+i\xi)(k_{1}+i\xi)}, \\
\sum\limits_{n=0}^{+\infty}\alpha^{n} & = &
    \alpha^{0} \displaystyle\frac{1-R_{1}^{+}R_{0}^{-}}{F_{sub}} =
    \displaystyle\frac{2k \biggl(1 + \displaystyle\frac{k_{1}-i\xi}
    {k_{1}+i\xi}e^{2ik_{1}R_{1}}\biggr) e^{-(\xi+ik)R_{2}}}
    {F_{sub} (k+i\xi)}, \\
\sum\limits_{n=0}^{+\infty}\beta^{n} & = &
    \displaystyle\frac{\sum\limits_{n=0}^{+\infty}\alpha^{n} -
    T_{2}^{-}}{R_{2}^{+}} =
    \alpha^{0} \displaystyle\frac{R_{1}^{-}(1 - R_{1}^{+}R_{0}^{-})
    + T_{1}^{-}R_{0}^{-}T_{1}^{+}}{F_{sub}} = \\
 & = & \displaystyle\frac{2k
     \biggl(\displaystyle\frac{i\xi-k_{1}}{i\xi+k_{1}} -
     e^{2ik_{1}R_{1}}\biggr) e^{\xi(2R_{1}-R_{2}) - ikR_{2}} }
     {F_{sub} (k+i\xi) },
\end{array}
\label{eq4_6}
\end{equation}                                  
where
\begin{equation}
\begin{array}{lcl}
F_{sub} & = & (1 - R_{1}^{+}R_{0}^{-})(1 - R_{2}^{+}R_{1}^{-}) -
        R_{2}^{+}T_{1}^{-}R_{0}^{-}T_{1}^{+} =
        1 + \displaystyle\frac{k_{1}-i\xi}{k_{1}+i\xi}
        e^{2ik_{1}R_{1}} - \\
& - &   \displaystyle\frac{(k-i\xi)(k_{1}-i\xi)}{(k+i\xi)(k_{1}+i\xi)}
        e^{-2\xi (R_{2} - R_{1})} -
        \displaystyle\frac{k-i\xi}{k+i\xi}
        e^{-2\xi (R_{2}-R_{1}) + 2ik_{1}R_{1}}.
\end{array}
\label{eq4_7}
\end{equation}                                  
\begin{equation}
\begin{array}{ll}
T_{2}^{-} = \alpha^{0} = 
            \displaystyle\frac{2k}{k+i\xi} e^{-(\xi+ik)R_{2}}, &
R_{2}^{-} = S^{0} = 
            \displaystyle\frac{-i\xi+k}{i\xi+k}e^{-2ikR_{2}}, \\
T_{1}^{-} = \displaystyle\frac{A^{n}}{\alpha^{n}} = 
            \displaystyle\frac{2i\xi}{i\xi+k_{1}} e^{(\xi+ik_{1})R_{1}}, &
R_{1}^{-} = \displaystyle\frac{\beta^{n}}{\alpha^{n}} = 
            \displaystyle\frac{i\xi-k_{1}}{i\xi+k_{1}} e^{2\xi R_{1}}, \\
T_{0}^{-} = 0, & 
R_{0}^{-} = 1, \\
T_{1}^{+} = \displaystyle\frac{\beta^{n+1}}{A^{n}} = -
            \displaystyle\frac{2k_{1}}{i\xi+k_{1}} e^{(\xi+ik_{1})R_{1}}, &
R_{1}^{+} = \displaystyle\frac{A^{n+1}}{A^{n}} = 
            \displaystyle\frac{i\xi-k_{1}}{i\xi+k_{1}} e^{2ik_{1}R_{1}}, \\
T_{2}^{+} = \displaystyle\frac{S^{n+1}}{\beta^{n}} = 
            \displaystyle\frac{2i\xi}{i\xi+k} e^{-(\xi+ik)R_{2}}, &
R_{2}^{+} = \displaystyle\frac{\alpha^{n+1}}{\beta^{n}} = 
            \displaystyle\frac{i\xi-k}{i\xi+k} e^{-2\xi R_{2}}, 
\end{array}
\label{eq4_8}
\end{equation}                                  
where the coefficients $T_{i}^{\pm}$ and $R_{i}^{\pm}$ are defined
in relation to the boundary with number $i$ ($i=0$ for $r=0$, $i=1$
for $r=R_{1}$ and $i=2$ for $r=R_{2}$). They can be calculated using
of the recurrence relations between the coefficients $S^{n}$, $A^{n}$,
$\alpha^{n}$ and $\beta^{n}$.

Now we consider the incident, transmitted and reflected w.p. in
relation to the barrier as a whole. Defining them for the region III,
one can write:
\begin{equation}
\begin{array}{lcl}
\chi_{inc}(r, t) & = & \int\limits_{0}^{+\infty} g(E - \bar{E})
        \theta(V_{1} - E) e^{-ikr -iEt/\hbar} dE, \\
\chi_{tr}(r, t)  & = & \int\limits_{0}^{+\infty} g(E - \bar{E})
        \theta(V_{1} - E) S_{tr} e^{ikr -iEt/\hbar} dE, \\
\chi_{ref}(r, t) & = & \int\limits_{0}^{+\infty} g(E - \bar{E})
        \theta(V_{1} - E) S_{ref} e^{ikr -iEt/\hbar} dE,
\end{array}
\label{eq4_9}
\end{equation}                                  
where
\begin{equation}
\begin{array}{lcl}
S_{tr}  & = & \sum\limits_{n=1}^{+\infty}S^{n}, \\
S_{ref} & = & S^{0}, \\
S       & = & S_{tr} + S_{ref}.
\end{array}
\label{eq4_10}
\end{equation}                                  

The expression $S$ represents a diagonal element of scattering
matrix corresponding to the orbital moment $l=0$. Thus, using
the method of multiple internal reflections it appears possible to
divide the S-matrix into two components corresponding to amplitudes
of stationary parts of the transmitted and reflected w.p. in relation
to the barrier as a whole. This property having physical sense, is
obtained for the first time.

The expressions for coefficients $S^{n}$, $A^{n}$, $\alpha^{n}$ and
$\beta^{n}$ for each step, the expression for the w.f. for each step,
the coefficients $T_{i}^{\pm}$ and $R_{i}^{\pm}$, the series of
the coefficients $S^{n}$, $A^{n}$, $\alpha^{n}$ and $\beta^{n}$ under
the substitution (\ref{eq2_14}) (and also at replacement of a sign
before argument for $\theta$-function at a consideration of the
non-stationary w.p.) transform into the corresponding expressions
for a solution of a problem of a w.p. propagation above the barrier.
The series (\ref{eq4_6}) of the coefficients $S^{n}$, $A^{n}$, $\alpha^{n}$
and $\beta^{n}$ coincide with the corresponding coefficients $S$, $A$,
$\alpha$ and $\beta$ for Eq.\ (\ref{eq4_3}), calculated by stationary
methods.

\subsection{Tunneling and reflecting times in relation to the barrier}

One can determine an equation for a propagation of a maximum
of the incident, transmitted and reflected w.p. in relation to the
barrier for the spherically symmetric problem. For radial parts of
non-stationary w.f. one can write:
\begin{equation}
\displaystyle\frac{\partial}{\partial E} \mbox{arg } \chi_{inc}(r, t)   =
\displaystyle\frac{\partial}{\partial E} \mbox{arg } \chi_{tr}(r, t)    =
\displaystyle\frac{\partial}{\partial E} \mbox{arg } \chi_{ref}(r, t) =
const.
\label{eq4_11}
\end{equation}                                  

Let's consider the first step of the propagation of the w.p.. Let the w.p.
is incident in the region III upon the external boundary of the barrier
at point $r=R_{2}$ in a time moment $t_{inc}$. Using Eq.\ (\ref{eq4_11}),
we find the time moment $t_{ref}^{1}$ of leaving outside from this
boundary the reflected w.p. in the region  III:
\begin{equation}
t_{ref}^{1} = t_{inc} + 
        \displaystyle\frac{2mR_{2}}{\hbar k} +
        \hbar\displaystyle\frac{\partial \mbox{arg } S^{0}}{\partial E}. 
\label{eq4_12}
\end{equation}                                  

Similarly, for a time moment $t_{tr}^{n}$ of leaving outside from
the external boundary of the barrier the $n$-multiple transmitted w.p.
one can write:
\begin{equation}
t_{tr}^{n} = t_{inc} + 
        \displaystyle\frac{2mR_{2}}{\hbar k} +
        \hbar\displaystyle\frac{\partial \mbox{arg } S^{n}}{\partial E}. 
\label{eq4_13}
\end{equation}                                  

Using Eq.\ (\ref{eq4_11}) at point $r = R_{2}$, we find
times necessary for the penetration of the total w.p. through the
barrier (describing the tunneling of the particle through the
barrier) and for the reflection of the w.p. from the barrier
(describing the reflection of the particle from the barrier):
\begin{equation}
\begin{array}{lcl}
\tau_{tun}^{Ph} & = & t_{tr} - t_{inc} =
        \displaystyle\frac{2mR_{2}}{\hbar k} +
        \hbar\displaystyle\frac{\partial \mbox{arg } S_{tr}}{\partial E}, \\
\tau_{ref}^{Ph} & = & t_{ref} - t_{inc} =
        \displaystyle\frac{2mR_{2}}{\hbar k} +
        \hbar\displaystyle\frac{\partial \mbox{arg } S_{ref}}{\partial E}.
\end{array}
\label{eq4_14}
\end{equation}                                  

For the problem of the w.p. tunneling under the barrier we receive:
\begin{equation}
\begin{array}{lcl}
\tau_{tun}^{Ph} & = & \hbar\displaystyle\frac{\partial}{\partial E}
        \mbox{arg } \displaystyle\frac{i\xi-k_{1}-(i\xi+k_{1})e^{2ik_{1}R_{1}}}
        {(i\xi+k)^{2}(i\xi+k_{1}) F_{sub}}, \\
\tau_{ref}^{Ph} & = & \displaystyle\frac{2m}{\hbar\xi k}.
\end{array}
\label{eq4_15}
\end{equation}                                  

For the problem of the w.p. propagating above the barrier we write:
\begin{equation}
\begin{array}{lcl}
\tau_{tun}^{Ph} & = & \displaystyle\frac{2m(R_{2}-R_{1})}{\hbar k_{2}} +
        \hbar\displaystyle\frac{\partial}{\partial E}
        \mbox{arg } \displaystyle\frac{k_{2}-k_{1}-(k_{2}+k_{1})e^{2ik_{1}R_{1}}}
        {(k+k_{2})(k_{1}+k_{2})F_{above}}, \\
\tau_{ref}^{Ph} & = & 0,
\end{array}
\label{eq4_16}
\end{equation}                                  
where $F_{above}$ can be obtained from $F_{sub}$ using the
substitution (\ref{eq2_14}).

Let's consider a particle, which tunnels under a high enough and wide
barrier. Then for the time of the tunneling we obtain the following expression
(sequence of approaches: $\xi (R_{2} - R_{1}) \to +\infty$, $\xi \to
+\infty$, $R_{2} - R_{1} \to + \infty$):
\begin{equation}
\tau_{tun}^{Ph} = \displaystyle\frac{2m}{\hbar k \xi} +
        \displaystyle\frac{4mR_{1}\sin{2k_{1}R_{1}}(1-2\cos{2k_{1}R_{1}})}
        {\hbar \xi(1-\cos{2k_{1}R_{1}})}.
\label{eq4_17}
\end{equation}                                  
The tunneling time does not depend on a width of the barrier
(Hartman's and Fletcher's effect), but depends on $k_{1}$ and $R_{1}$.

\section{Tunneling of the particle through the spherically symmetric
barrier of a general view
\label{sec:level5}}

\subsection{The particle propagates above the barrier}

In study of nuclear processes when a tunneling of partiáles through
a barrier is investigated, in the most cases the barriers of more
complicated form than rectangular are used.
So, the spherically symmetric two-humb potential of Strutinski has an
enough important role in problems of fission and decai of nuclei.
A degree of an exactitude of a description of the nuclear
process depends on choice of a form of the potential. Therefore, we shall
consider, as far as it is possible to use the method of multiple
internal reflections for solving the spherically symmetrical problems
with the barrier of a general view.

Let's consider a particle propagating in a spherically symmetric
potential field, which radial part has a barrier.
Taking into account a behaviour of a radial part $V(r)$ of
the potential function in dependence on $r$, we divide the area of
its definition $r \in [0; +\infty[$ on $n$ regions.
In each region let's replace the potential function $V(r)$ by
a function most close describing $V(r)$ and for which an exact solution
of the stationary Schr\"{o}dinger equation exists (see Fig.\ \ref{fig4}).
Passing to the problem of the particle propagation in the field of
these approximated potential functions, we write the general solution
for stationary w.f. in the form (\ref{eq4_2}), where its radial part
can be written as
\begin{equation}
\chi(r) = \left\{
\begin{array}{lll}
A_{1} a_{1}(k, r) + B_{1} b_{1}(k, r), & \mbox{for }  0<r<R_{1},
        & \mbox{(region I)}, \\
A_{i} a_{i}(k, r) + B_{i} b_{i}(k, r), & \mbox{for }  R_{i-1}<r<R_{i},
        & \mbox{(region  i)}, \\
A_{n} a_{n}(k, r) + B_{n} b_{n}(k, r), & \mbox{for }  r>R_{n-1},
        & \mbox{(region N)},
\end{array} \right.
\label{eq5_1}
\end{equation}                                  
where $k = \frac{1}{\hbar} \sqrt{2mE}$,
      $a_{i}(k, r)$ and $b_{i}(k, r)$ are the partial solutions of
      the radial part of w.f. in region $i$,
      $A_{i}$ and $B_{i}$ are the normalization constants.

Let's find the transmission and reflection coefficients of particle
in relation to the barrier, and also the times necessary for transmission
and for reflection of the particle in relation to the barrier, using
the method of multiple internal reflections. To apply the method to
this problem, one can need to present the general stationary solution
of w.f. in each region in the sum of divergent and convergent vawes.

Using the Fourier transformation, one can write:
\begin{equation}
\begin{array}{lcl}
a_{i}(k, r) & = & a_{i}^{-}(k, r) + a_{i}^{+}(k, r), \\
b_{i}(k, r) & = & b_{i}^{-}(k, r) + b_{i}^{+}(k, r), 
\end{array}
\label{eq5_2}
\end{equation}                                  
where
\begin{equation}
\begin{array}{lcl}
a_{i}^{-}(k, r) & = &
        \displaystyle\frac{1}{\sqrt{2\pi}}
        \int\limits_{-\infty}^{0} dq
        \int\limits_{R_{i-1}}^{R_{i}} 
        a_{i}(k, r') e^{iq(r-r')} dr',  \\
a_{i}^{+}(k, r) & = &
        \displaystyle\frac{1}{\sqrt{2\pi}}
        \int\limits_{0}^{+\infty} dq
        \int\limits_{R_{i-1}}^{R_{i}} 
        a(k, r') e^{iq(r-r')} dr',      \\
b_{i}^{-}(k, r) & = &
        \displaystyle\frac{1}{\sqrt{2\pi}}
        \int\limits_{-\infty}^{0} dq
        \int\limits_{R_{i-1}}^{R_{i}} 
        b_{i}(k, r') e^{iq(r-r')} dr',  \\
b_{i}^{+}(k, r) & = &
        \displaystyle\frac{1}{\sqrt{2\pi}}
        \int\limits_{0}^{+\infty} dq
        \int\limits_{R_{i-1}}^{R_{i}} 
        b_{i}(k, r') e^{iq(r-r')} dr'.
\end{array}
\label{eq5_3}
\end{equation}                                  

Taking into account the transformation
\begin{equation}
\begin{array}{lcl}
c_{i}^{-}(k, r) & = & a_{i}^{-}(k, r) +
                  \displaystyle\frac{B_{i}}{A_{i}} b_{i}^{-}(k, r), \\
c_{i}^{+}(k, r) & = & \displaystyle\frac{A_{i}}{B_{i}} a_{i}^{+}(k, r) +
                  b_{i}^{+}(k, r)
\end{array}
\label{eq5_4}
\end{equation}                                  
one can write the general solusion (\ref{eq5_1}) as
\begin{equation}
\chi(r) = \left\{
\begin{array}{lll}
A_{1} c_{1}^{-}(k, r) + B_{1} c_{1}^{+}(k, r),
        & \mbox{for }  0<r<R_{1},
        & \mbox{(region I)},           \\
A_{i} c_{i}^{-}(k, r) + B_{i} c_{i}^{+}(k, r),
        & \mbox{for }  R_{i-1}<r<R_{i},
        & \mbox{(region i)},           \\
A_{n} c_{n}^{-}(k, r) + B_{n} c_{n}^{+}(k, r),
        & \mbox{for }  r>R_{n-1},
        & \mbox{(region N)},
\end{array} \right.
\label{eq5_5}
\end{equation}                                  

In result, the general solution in every region $i$ is represented
as the sum of convergent vawes $c_{i}^{-}(r)$ and divergent vawes
$c_{i}^{+}(r)$ (so, in case of a rectangular barrier in the region
$i$ such expressions
equal to $e^{-ik_{i}r}$ and $e^{ik_{i}r}$, accordingly).
On the basis of these expressions using Eq.\ (\ref{eq4_4})
one can construct the non-stationary convergent
and divergent w.p.. Writing the genegal solution for w.f. in
every region in the form of linear combination of convergent
and divergent w.p., one can apply the method of multiple
internal reflections for solving the problem.

At first we study the case, when the general stationary solution
for w.f. in every region can be written uniquely as sum of
convergent and divergent vawes $c_{i}^{\pm}(r)$. Then using of
the method of multiple internal reflections for solving the
problem, one can find the incident, transmitted and reflected
w.p. in relation to the barrier at whole, and total w.p. in
every region. It is enough convenient to use the coefficients
$T_{i}^{\pm}$ and $R_{i}^{\pm}$ (as in the spherically
symmetric problem with rectangular barrier). We define these
coefficients in relation to the boundary with the number $i$
by such way (for the step $j$):
\begin{equation}
\begin{array}{lcllcl}
A_{i}^{j}     & = & T_{i}^{-} A_{i+1}^{j}; &
B_{i+1}^{j+1} & = & R_{i}^{-} A_{i+1}^{j}; \\
B_{i+1}^{j}   & = & T_{i}^{+} B_{i}^{j};   &
A_{i}^{j+1}   & = & R_{i}^{+} B_{i}^{j}.
\end{array}
\label{eq5_6}
\end{equation}                                  

One can calculate these coefficients at consideration of first $2n+1$
steps:
\begin{equation}
\begin{array}{ll}
T_{0}^{-} = 0; &
R_{0}^{-} = - \displaystyle\frac{c_{1}^{-}(0)}{c_{1}^{+}(0)} =
            \displaystyle\frac{B_{1}}{A_{1}},
\end{array}
\label{eq5_7}
\end{equation}                                  
\begin{equation}
\begin{array}{lcl}
T_{i}^{-} & = & \displaystyle\frac{
        \displaystyle\frac{\partial c_{i+1}^{-}(r)}{\partial r}
        c_{i+1}^{+}(r) -
        c_{i+1}^{-}(r)
        \displaystyle\frac{\partial c_{i+1}^{+}(r)}{\partial r}}
        {\displaystyle\frac{\partial c_{i}^{-}(r)}{\partial r}
        c_{i+1}^{+}(r) -
        c_{i}^{-}(r)
        \displaystyle\frac{\partial c_{i+1}^{+}(r)}{\partial r}}
        \Biggl|_{r = R_{i}}, \\
R_{i}^{-} & = & \displaystyle\frac{
        \displaystyle\frac{\partial c_{i+1}^{-}(r)}{\partial r}
        c_{i}^{-}(r) -
        c_{i+1}^{-}(r)
        \displaystyle\frac{\partial c_{i}^{-}(r)}{\partial r}}
        {\displaystyle\frac{\partial c_{i}^{-}(r)}{\partial r}
        c_{i+1}^{+}(r) -
        c_{i}^{-}(r)
        \displaystyle\frac{\partial c_{i+1}^{+}(r)}{\partial r}}
        \Biggl|_{r = R_{i}}, \\
T_{i}^{+} & = & \displaystyle\frac{
        \displaystyle\frac{\partial c_{i}^{+}(r)}{\partial r}
        c_{i}^{-}(r) -
        c_{i}^{+}(r)
        \displaystyle\frac{\partial c_{i+1}^{-}(r)}{\partial r}}
        {\displaystyle\frac{\partial c_{i+1}^{+}(r)}{\partial r}
        c_{i}^{-}(r) -
        c_{i+1}^{+}(r)
        \displaystyle\frac{\partial c_{i}^{-}(r)}{\partial r}}
        \Biggl|_{r = R_{i}}, \\
R_{i}^{+} & = & \displaystyle\frac{
        \displaystyle\frac{\partial c_{i}^{+}(r)}{\partial r}
        c_{i+1}^{+}(r) -
        c_{i}^{+}(r)
        \displaystyle\frac{\partial c_{i+1}^{-}(r)}{\partial r}}
        {\displaystyle\frac{\partial c_{i+1}^{+}(r)}{\partial r}
        c_{i+1}^{+}(r) -
        c_{i+1}^{+}(r)
        \displaystyle\frac{\partial c_{i}^{-}(r)}{\partial r}}
        \Biggl|_{r = R_{i}}.
\end{array}
\label{eq5_8}
\end{equation}                                  

Further using the method of multiple internal reflections, one can
calculate the incident, transmitted  and reflected w.p. in relation
to the barrier. On the basis of these w.p. one can find the transmission
and reflection coefficients and also the transmission and reflection
times in relation to the barrier. Thus the transmitted and reflected
w.p. can be written through $S_{tr}$ and $S_{ref}$, which sum is the
diagonal element of the scattering matrix $S$ at the orbital moment $l$.

The value $A_{n}$ can be obtained from the normalization condition:
\begin{equation}
A_{n} = \Biggl(\int\limits_{R_{n-1}}^{+\infty} |c_{n}(k, r)|^{2} dr
        \Biggr)^{-1/2}.
\label{eq5_9}
\end{equation}                                  

Now we study the case, when the partial solutions of w.f. in some
regions are not the convergent and divergent vawes. In representations
(\ref{eq5_4}) and (\ref{eq5_5}) one can need to know the values
$B_{i} / A_{i}$. In this case at solving the problem we consider
first $2n+1$ steps. Let the general solution for w.f. in first
region be expressed through $a_{1}$ and $b_{1}$. Analyzing the
reflection of w.p. from point $r = 0$, one can obtain:
\begin{equation}
\begin{array}{ll}
T_{0}^{-} = 0; &
R_{0}^{-} = - \displaystyle\frac{a_{1}(0)}{b_{1}(0)} =
            \displaystyle\frac{B_{1}}{A_{1}}.
\end{array}
\label{eq5_10}
\end{equation}                                  

If the w.f. in the first region is determined through $c_{1}^{\pm}$
uniquely, then it is need to use Eq.\ (\ref{eq5_7}) instead of
Eq.\ (\ref{eq5_10}). Calculating the value $B_{1} / A_{1}$,
then one can find the functions $c_{1}^{\pm}$. Using the
continuity condition for w.f. and its derivative in all
boundaries between regions, one can find the recurrent relation
for values $B_{i} / A_{i}$:
\begin{equation}
\begin{array}{l}
\displaystyle\frac{B_{i+1}}{A_{i+1}} =
        \displaystyle\frac{
        f_{i}(r)
        \displaystyle\frac{\partial a_{i+1}(r)}{\partial r} -
        \displaystyle\frac{\partial f_{i}(r)}{\partial r}
        a_{i+1}(r)}
        {\displaystyle\frac{\partial f_{i}(r)}{\partial r}
        b_{i+1}(r) -
        f_{i}(r)
        \displaystyle\frac{\partial b_{i+1}(r)}{\partial r}}
        \Biggl|_{r = R_{i}}, \\
f_{i}(r) =
        a_{i}(r) +
        b_{i}(r) \displaystyle\frac{B_{i}}{A_{i}} \Biggl|_{r = R_{i}} =
        c_{i}^{-}(r) +
        c_{i}^{+} \displaystyle\frac{B_{i}}{A_{i}} \Biggl|_{r = R_{i}}.
\end{array}
\label{eq5_11}
\end{equation}                                  

Having the values $B_{i} / A_{i}$, we obtain the convergent
and divergent vawes $c_{i}^{\pm}(r)$ in every region. Then the
solution of problem is fulfilled as in the previous case.

Applying the approach considered above for the solution of problem of
particle propagation, when the potential $V(r)$ is defined only in two
regions ($n=2$), one can find the incident, transmitted and reflected
w.p. in relation to the barrier using Eq.\ (\ref{eq4_4}) for
above-barrier region, where the radial parts from corresponding
stationary w.f. have the form (at $r > R_{1}$)
\begin{equation}
\begin{array}{lcll}
\chi_{inc}(r) & = & A_{2}  c_{2}^{-}(k, r), \\
\chi_{tr}(r)  & = & S_{tr} c_{2}^{+}(k, r) =
        A_{2} \displaystyle\frac{T_{1}^{-}T_{1}^{+} R_{0}^{-}}
        {1-R_{1}^{+}R_{0}^{-}} c_{2}^{+}(k, r), \\
\chi_{ref}(r) & = & S_{ref} c_{2}^{+}(k, r) =
        A_{2} R_{1}^{-} c_{2}^{+}(k, r).
\end{array}
\label{eq5_12}
\end{equation}                                  

We find the coefficients $T_{i}^{\pm}$ and $R_{i}^{\pm}$ from Eq.\
(\ref{eq5_8}). Using Eq.\ (\ref{eq4_11}) for external boundary, one can
obtain the times necessary for transmission and for reflection of particle
in relation to the barrier. In result, we receive:
\begin{equation}
\begin{array}{lcl}
\tau_{tun} & = & \hbar\displaystyle\frac{\partial}{\partial E}
                 \Biggl( \mbox{arg } \biggl(\sum\limits_{n=1}^{+\infty} 
                 B_{2}^{i} c_{2}^{+}(k, R_{1}) \biggr) -
                 \mbox{arg } A_{2} c_{2}^{-}(k, R_{1}) \Biggr) = \\
           & = & \hbar\displaystyle\frac{\partial}{\partial E}
                 \mbox{arg } \displaystyle\frac{T_{1}^{-}T_{1}^{+}R_{0}^{-}}
                 {1-R_{1}^{+}R_{0}^{-}} + \Delta \tau;          \\
\tau_{ref} & = & \hbar\displaystyle\frac{\partial}{\partial E}
                 \Biggl( \mbox{arg } \biggl(B_{2}^{0}
                 c_{2}^{+}(k, R_{1}) \biggr) -
                 \mbox{arg } A_{2} c_{2}^{-}(k, R_{1}) \Biggr) = \\
           & = & \hbar\displaystyle\frac{\partial}{\partial E}
                 \mbox{arg } R_{1}^{-} + \Delta \tau;           \\
\Delta\tau & = & \hbar\displaystyle\frac{\partial}{\partial E}
                 \mbox{arg } \displaystyle\frac{
                 c_{2}^{+}(R_{1})}{c_{2}^{-}(R_{1})}.
\end{array}
\label{eq5_13}
\end{equation}                                  

For the problem solution when potential $V(r)$ is defined on three
regions ($n=3$), the expressions for radial parts of stationary w.f.,
describing the incident, transmitted and reflected w.p. in relation to
the barrier, look like (at $r > R_{2}$)
\begin{equation}
\begin{array}{lcll}
\chi_{inc}(r) & = & A_{3}  c_{3}^{-}(k, r), \\
\chi_{tr}(r)  & = & S_{tr} c_{3}^{+}(k, r) =
        A_{3} \displaystyle\frac{T_{2}^{-}T_{2}^{+} (R_{1}^{-}(1 -
        R_{1}^{+}R_{0}^{-}) + T_{1}^{-}R_{0}^{-}T_{1}^{+})}
        {(1-R_{1}^{+}R_{0}^{-}) (1-R_{2}^{+}R_{1}^{-}) -
        R_{2}^{+}T_{1}^{-}R_{0}^{-}T_{1}^{+}} c_{3}^{+}(k, r) \\
\chi_{ref}(r) & = & S_{ref} c_{3}^{+}(k, r) =
        A_{3} R_{2}^{-} c_{3}^{+}(k, r).
\end{array}
\label{eq5_14}
\end{equation}                                  

The transmission and reflection times of particle in relation to the
barrier has the form (they are calculated at $r=R_{2}$)
\begin{equation}
\begin{array}{lcl}
\tau_{tun} & = & 
        \hbar\displaystyle\frac{\partial}{\partial E}
        \mbox{arg } \displaystyle\frac{T_{2}^{-}T_{2}^{+} (R_{1}^{-}(1 -
        R_{1}^{+}R_{0}^{-}) + T_{1}^{-}R_{0}^{-}T_{1}^{+})}
        {(1-R_{1}^{+}R_{0}^{-}) (1-R_{2}^{+}R_{1}^{-}) -
        R_{2}^{+}T_{1}^{-}R_{0}^{-}T_{1}^{+}} +
        \Delta\tau; \\
\tau_{ref} & = & 
        \hbar\displaystyle\frac{\partial}{\partial E}
        \mbox{arg } R_{2}^{-} + \Delta\tau; \\
\Delta\tau & = & 
        \hbar\displaystyle\frac{\partial}{\partial E}
        \mbox{arg } \displaystyle\frac{c_{3}^{+}(R_{2})}{c_{3}^{-}(R_{2})}.
\end{array}
\label{eq5_15}
\end{equation}                                  

\vspace *{\baselineskip}

As an examples of method application we consider two problems.

The particle propagates above the barrier of form (see Fig.\ \ref{fig5})
\begin{equation}
V(r) = \left\{
\begin{array}{cll}
   -V_{0},      & \mbox{for } 0<r<R_{1}, & \mbox{(region I)}; \\
   \displaystyle\frac{\gamma}{r},
                & \mbox{for } r>R_{1},   & \mbox{(region II)}.
\end{array} \right.
\label{eq5_16}
\end{equation}                                  
We consider the case $l \neq 0$. One can obtain the incident,
transmitted and reflected w.p. in relation to the barrier from
Eqs.\ (\ref{eq5_12}) and (\ref{eq4_4}), taking into account the
sign before argument of $\theta$-function for above-barrier energies,
and transmission and reflection times from Eq.\ (\ref{eq5_13}).
At consideration the first three steps of w.p. propagation along the
barrier we find the coefficients $T_{i}^{\pm}$ and $R_{i}^{\pm}$ using
Eqs.\ (\ref{eq5_8}) for $n=2$. In the solutions one can need to
fulfil the substitution
\begin{equation}
\begin{array}{lcllcl}
c_{1}^{-}(k, r) & = & \chi_{k_{1}l}^{-}(r),                         &
c_{2}^{-}(k, r) & = & G_{l}(\eta, \rho) - iF_{l}(\eta, \rho),       \\
c_{1}^{+}(k, r) & = & \chi_{k_{1}l}^{+}(r),                         &
c_{2}^{+}(k, r) & = & G_{l}(\eta, \rho) + iF_{l}(\eta, \rho),
\end{array}
\label{eq5_17}
\end{equation}                                  
where
\begin{equation}
\begin{array}{lcllcl}
k_{1} & = & \frac{1}{\hbar}\sqrt{2m(E + V_{0})},        \\
\eta  & = & \displaystyle\frac{\mu\nu k}{\hbar^{2}},    \\
\rho  & = & k(r),                                       \\
\chi_{k_{1}l}^{\pm}(r) & = & \pm i\sqrt{\displaystyle\frac{\pi
            k_{1}r}{2}} H_{l+1/2}^{(1, 2)}(k_{1}r), 
\end{array}
\label{eq5_18}
\end{equation}                                  
$H_{l}^{(1, 2)} (r)$ is the function of Hankel of 1-st and 2 sort, \newline
$G_{l} (\eta, \rho)$ and $F_{l} (\eta, \rho)$ are the irregular and regular
Coulomb functions \cite{Abramowitz}.
The normalization constant $A_{2}$ can be obtained from Eq.\ (\ref{eq5_9}).

\vspace *{\baselineskip}

Now we consider another problem when the particle propagates above
the barrier of following form:
%
\begin{equation}
V(r) = \left\{
\begin{array}{cll}
   \alpha r^{2} - V_{0},
                & \mbox{for } 0<r<R_{1}, & \mbox{(region I)}; \\
   \displaystyle\frac{\gamma}{r},
                & \mbox{for } r>R_{1},   & \mbox{(region II)}.
\end{array} \right.
\label{eq5_19}
\end{equation}                                  

Let's study the case $l = 0$. In the beginning we consider region I.
The partial solutions for the radial part of stationary w.f. are the
parabolic cylinder functions \cite{Abramowitz}:
$D_{\nu} (\pm gr)$ and $D_{\nu} (\pm igr)$, where
$g = (8\alpha\mu / \hbar^{2})^{1/4}$.
For the description of above-barrier motion of particle we choose the
first two solutions $D_{\nu} (\pm gr)$, which are independent if $\nu$
is non-integer. (Note that one can use the Whitteker's functions as
such two independent solutions \cite{Abramowitz}. But these two functions
can be presented in the form of linear combination of the parabolic
cylinder functions $D_{\nu} (\pm gr)$.)
Each of partial solutions can be presented in the form of sum
of convergent and divergent waves:
\begin{equation}
\begin{array}{lcl}
D_{\nu}(\pm gr) & = & D_{\nu}^{-}(\pm gr) + D_{\nu}^{+}(\pm gr), \\
D_{\nu}(\pm gr)^{-} & = &
        \displaystyle\frac{1}{\sqrt{2\pi}}
        \int\limits_{-\infty}^{0} dq
        \int\limits_{0}^{R_{1}}
        D_{\nu}(\pm gr') e^{iq(r-r')} dr', \\
D_{\nu}^{+}(\pm gr) & = &
        \displaystyle\frac{1}{\sqrt{2\pi}}
        \int\limits_{0}^{+\infty} dq
        \int\limits_{0}^{R_{1}}
        D_{\nu}(\pm gr') e^{iq(r-r')} dr'.
\end{array}
\label{eq5_20}
\end{equation}                                  

Using such w.f., one can apply the method of multiple internal reflections
to the solution of problem. In result, we find the incident, transmitted
and reflected w.p. in relation to the barrier from Eqs.\ (\ref{eq5_12}) and
(\ref{eq4_4}), taking into account the sign before argument of
$\theta$-function for above-barrier energies, and transmission and
reflection times from Eq.\ (\ref{eq5_13}).
The coefficients $T_{i}^{\pm}$ and $R_{i}^{\pm}$ can be obtained from Eqs.\
(\ref{eq5_8}) and (\ref{eq5_10}) for $n=2$ at substitution
\begin{equation}
\begin{array}{lcllcl}
a_{1}(k, r)     & = & D_{\nu}(k, r),                            &
c_{2}^{-}(k, r) & = & G_{0}(\eta, \rho) - iF_{0}(\eta, \rho),   \\
b_{1}(k, r) & = & D_{\nu}(-k, r),                               &
c_{2}^{+}(k, r) & = & G_{0}(\eta, \rho) + iF_{0}(\eta, \rho),
\end{array}
\label{eq5_21}
\end{equation}                                  
where
$\eta$ and $\rho$ are defined in Eq.\ (\ref{eq5_18}),
$G_{0} (\eta, \rho)$ and $F_{0} (\eta, \rho)$ are the irregular and regular
Coulomb functions at $l=0$.

\subsection{The particle tunnels under the barrier}

Now we consider the problem of tunneling of particle under the barrier
of spherically symmetric potential field. And the radial part of this
barrier has a general view (see Fig.\ \ref{fig4}).

Dividing the range of definition $r \in [0; +\infty[$ for potential
$V(r)$ on $n$ regions, on each of them we approximate $V(r)$ by function
most close to it, for which there are the general solutions of w.f.
for stationary Schr\"{o}dinger equation. We divide the whole range
so that the processes of sub-barrier tunneling and above-barrier
propagation laid in the different regions.

For regions, in which the energy levels considered by us, lay above
the potential function $V(r)$ (the particle propagates above the
potential), the stationary solution for w.f. is represented as Eq.\
(\ref{eq5_5}) (if necessary using the transformations (\ref{eq5_2}),
(\ref{eq5_3}) and (\ref{eq5_4})).

For regions, in which the viewed energy levels lay under the potential
function $V(r)$ (the particle tunnels under the potential), in the
beginning we find the general solution for stationary w.f., assuming,
that the energy levels lay above the potential function. One can need
to present the general solution for w.f. as Eq.\ (\ref{eq5_5}), separate
the components corresponding to fluxes, directed to the opposite sides.
Everywhere in expressions for w.f., where the property
\begin{equation}
\begin{array}{ll}
|E - V_{l}(r)| = E - V_{l}(r), & \mbox{¯à¨ } E > V_{l}(r),
\end{array}
\label{eq5_22}
\end{equation}                                  
is used, one can need to redefine this expression for $E < V_{l}(r)$,
having changed the sign.
So, in case of constant potential in dependence on $r$ we obtain the
Eq.\ (\ref{eq2_14}). Such substitution gives the following property: the
resultant expressions for w.p. and also for stationary and
non-stationary w.f. for the problem of tunneling of a particle under
the barrier are the analytic continuation of the relevant expressions for
a similar problem, when the particle propagates above the barrier.

Having defined the expressions for stationary w.f. by such way, one can
construct the relevant them w.p. on each region and apply the method
of multiple internal reflections to solution of the problem. The further
approach for obtaining the resultant expressions for incident,
transmitted and reflected w.p. in relation to the barrier and also
the times of tunneling and reflection differs by nothing from the approach
for the problem solution in the above-barrier case.

As an example, we consider the problem of tunneling of particle under
the barrier (\ref{eq5_16}) (see Fig.\ \ref{fig5}). We consider the case
$l \neq 0$. We divide the region II on two at point $r = R_{2}$, which
defines by requirement $E = V_{l}(R_{2})$. One can find the incident,
transmitted and reflected w.p. in relation to the barrier from Eqs.\
(\ref{eq4_4}) and (\ref{eq5_14}), and the times of tunneling and reflection
from Eq.\ (\ref{eq5_15}). Analysing the first 5 steps of w.p. propagation
along the barrier, we find the coefficients $T_{i}^{\pm}$ and $R_{i}^{\pm}$
using the Eqs.\ (\ref{eq5_8}) for $n=3$. In these expressions on can need
to fulfil the substitution
\begin{equation}
\begin{array}{l}
\begin{array}{l}
c_{1}^{-}(k, r) = \chi_{k_{1}l}^{-}(r),                             \\
c_{1}^{+}(k, r) = \chi_{k_{1}l}^{+}(r),
\end{array}                                                     \\
\left. \begin{array}{l}
   c_{2}^{-}(k, r) = G_{l}(\eta, \rho) - iF_{l}(\eta, \rho)         \\
   c_{2}^{+}(k, r) = G_{l}(\eta, \rho) + iF_{l}(\eta, \rho)
   \end{array} \right\}                                         
   \mbox{for } E < \displaystyle\frac{\gamma}{r},               \\
\left. \begin{array}{l}
   c_{3}^{-}(k, r) = G_{l}(\eta, \rho) - iF_{l}(\eta, \rho)         \\
   c_{3}^{+}(k, r) = G_{l}(\eta, \rho) + iF_{l}(\eta, \rho)
   \end{array} \right\}
   \mbox{for } E > \displaystyle\frac{\gamma}{r},
\end{array}
\label{eq5_23}
\end{equation}                                  
where $k$, $k_{1}$, $\eta$, $\rho$, and also $\chi_{k_{1}l}^{\pm}(r)$,
$G_{l}(\eta, \rho)$ and $F_{l}(\eta, \rho)$ are defined earlier.

\section {Evolution of photon tunneling through one-dimensional
undersized rectangular waveguide
\label{sec:level6}}

We use the analogy between photon and particle 1D
propagation and tunneling which consists not only in the formal
mathematical analogy between the solutions of the time-dependent
Schr\"{o}dinger equation for nonrelativistic particles and of the
time-dependent Helmholtz equation for electromagnetic waves but also
in the similarity of the probabilistic interpretation of the wave
function for a particle and of a an electromagnetic wave packet
being the wave function for a single photon \cite{Olkhovsky.1997.Trieste}.
For a hollow rectangular waveguide with variable section (like that
used in the Cologne experiment \cite{Enders.1992.JPGCE},
see Fig.\ \ref{fig6}).
The time-dependent wave equation for
${\bf A}$, ${\bf E}$, ${\bf H}$ (${\bf A}$ is the vector potential with
the subsidiary gauge condition $\mbox{div \bf A} = 0$,
${\bf E} = -(1/c) \partial {\bf A} / \partial t$ is the electric field
strength, ${\bf H} = \mbox{rot }{\bf A}$ is the magnetic field
strength) is
\begin{equation}
\Delta {\bf A} - \displaystyle\frac{1}{c^{2}}
\displaystyle\frac{\partial^{2} {\bf A}}{\partial t^{2}} = 0.
\label{eq_phot1}
\end{equation}                                  
For boundary conditions (see, for instance, \cite{Olkhovsky.1997.Trieste})
\begin{equation}
\begin{array}{ll}
  E_{y} = 0 & \mbox{ for } z = 0 \mbox{ and } z = a,      \\
  E_{z} = 0 & \mbox{ for } y = 0 \mbox{ and } y = b
\end{array}
\label{eq_phot2}
\end{equation}                                  
the solution of the Eq.\ (\ref{eq_phot2}) can be represented as a
superposition of the following monochromatic waves:
\begin{equation}
\begin{array}{lcl}
  E_{x} & = & 0, \\
  E_{y}^{\pm} & = & E_{0} \sin{(k_{z} z)} \cos{(k_{y} y)}
                    \exp{[i(wt \pm \gamma x)]}, \\
  E_{z}^{\pm} & = & -E_{0} (k_{y} / k_{z}) \cos{(k_{z} z)} \sin{(k_{y} y)}
                    \exp{[i(wt \pm \gamma x)]},
\end{array}
\label{eq_phot3}
\end{equation}                                  
where
$k_{z}^{2} + k_{y}^{2} + \gamma^{2} = w^{2} /
        c^{2} = (2\pi / \lambda)^{2}$,
$k_{z} = m\pi/a$,
$k_{y} = n\pi/b$,
$m$ and $n$ are the integer numbers (for definiteness we have chosen
the TE-waves). Thus,
\begin{equation}
\begin{array}{ll}
  \gamma = 2\pi [(1/\lambda)^{2} - (1/\lambda_{c})^{2}]^{1/2}, &
  (1/\lambda_{c})^{2} = (m/2a)^{2} + (n/2b)^{2},
\end{array}
\label{eq_phot4}
\end{equation}                                  
where $\gamma$ is real ($\gamma = Re \gamma$) if $\lambda <
\lambda_{c}$ and $\gamma$ is imaginary ($\gamma = i\xi_{em}$) if
$\lambda > \lambda_{c}$. Similar expressions for $\gamma$ were
obtained for TH-waves \cite{Olkhovsky.1997.Trieste}.

Generally the non-stationary solution of Eq.\ (\ref{eq_phot1})
can be written
as a wave packet constructed on the basis of monochromatic solutions
(\ref{eq_phot3}), similarly to the solution of the time-dependent
Schr\"{o}dinger equation for nonrelativistic particles in the form of a
wave packet constructed from monochromatic terms (for the problem of
particle propagating above the 1D rectangular barrier). Moreover, in
the representation of primary quantization the
probabilistic single-photon wave function is usually described by
a wave packet (for instance, see
\cite{Olkhovsky.1997.Trieste,Olkhovsky.1997.JPGCE} and the
relevant references therein) like
\begin{equation}
{\bf A} ({\bf r}, t) =
        \int\limits_{k_{0}>0} \displaystyle\frac{d^{3}k}{k_{0}}
        {\bf K}({\bf k}) A({\bf k}, {\bf r}) e^{-ik_{0}t},
\label{eq_phot5}
\end{equation}                                  
where $A ({\bf k}, {\bf r}) = \exp{(i{\bf kr})}$ for propagation
in vacuum and $A ({\bf k}, {\bf r}) = \varphi(x) \exp(ik_{y}y$ 
$+ ik_{z}z)$ with
\begin{equation}
\varphi(x) = \left \{
\begin{array}{ll}
   e^{ik_{x}x} + a_{R} e^{-ik{x}x},                & \mbox{ region I},   \\
   \alpha e^{-\xi_{em} x} + \beta e^{\xi_{em} x},  & \mbox{ region II},  \\
   a_{T} e^{ik_{x}x},                              & \mbox{ region III}
\end{array} \right.
\label{eq_phot6}
\end{equation}                                  
for propagation in the waveguide (Fig.\ \ref{fig6}). Here,
${\bf r} = (x, y, z)$,
${\bf K}({\bf k}) = \sum\limits_{i=1}^{2} K_{i}({\bf k})
{\bf e}_{i}({\bf k})$,
${\bf e}_{i} {\bf e}_{j} = \delta_{ij}$,
${\bf e}_{i}({\bf k}) {\bf k} = 0$,
$i, j = 1, 2$ (or $y$, $z$ if ${\bf kr} = k_{x}x$),
$k_{0} = w/c = \varepsilon/\hbar c$, $|{\bf k}| = k_{0}$,
$K_{i}({\bf k})$ is the amplitude for the photon with impulse
${\bf k}$ and polarization $i$, and $|K_{i}({\bf k})|^{2}{\bf dk}$ is
then proportional to the probability that the photon has the impulse
between ${\bf k}$ and ${\bf k} + {\bf dk}$ in the polarization state
${\bf e}_{i}$.

Though it is not possible to localize photon in the direction of its
polarization, nevertheless, in a certain sense, for the one-dimensional
propagation it is possible to use the space-time probabilistic
interpretation of Eq.\ (\ref{eq_phot5}) along axis $x$
(the propagation direction) \cite{Olkhovsky.1997.Trieste}.
It can be realized from the following. Usually one uses not the
probability density and probability flux density with the corresponding
continuity equation directly but the energy density $s_{0}$ and the
energy flux density $s_{x}$ (although in general they represent
components of not a 4-dimensional vector but the energy-momentum
tensor) with the corresponding continuity equation
\cite{Olkhovsky.1997.Trieste}
which we write in the two-dimensional (spatially one-dimensional)
form:
\begin{equation}
  \displaystyle\frac{\partial s_{0}}{\partial t} +
  \displaystyle\frac{\partial s_{x}}{\partial x} = 0,
\label{eq_phot7}
\end{equation}                                  
where
\begin{equation}
\begin{array}{ll}
   s_{0} = \displaystyle\frac{{\bf E}{\bf E} + {\bf H}{\bf H}}{8\pi}, &
   s_{x} = \displaystyle\frac{c Re [{\bf E}{\bf H}]_{x}}{2\pi}
\end{array}
\label{eq_phot8}
\end{equation}                                  
and axis $x$ is directed along the motion direction (the mean impulse)
of the wave packet (\ref{eq_phot5}).
Note, that for the spatially one-dimensional
propagation the energy-momentum tensor of the electromagnetic field
reduces to the two-component quantity --- to the scalar term $s_{0}$ and
1-dimensional vector term $s_{x}$ for which continuity equation
(\ref{eq_phot7}) is Lorentz-invariant.
Then, as a normalization condition one chooses the
equality of the spatial integrals of $s_{0}$ and $s_{x}$ to the mean
photon energy and the mean photon impulse respectively or simply the
unit energy flux density $s_{x}$. With this,
we can define conventionally the probability density
\begin{equation}
  \rho_{em} dx = \displaystyle\frac{S_{0} dx}{\int S_{0} dx},   \;
  S_{0} = \int s_{0} dy dz
\label{eq_phot9}
\end{equation}                                  
for the photon to be found (localized) in the spatial interval ($x$,
$x+dx$) along axis $x$ at the moment $t$, and the flux probability
\begin{equation}
   J_{em, x} dt = \displaystyle\frac{S_{x} dt}{\int S_{x} dt},  \;
   S_{x} = \int s_{x} dydz,
\label{eq_phot10}
\end{equation}                                  
for the photon to propagate through point (plane) $x$ in the time
interval ($t$, $t+dt$), quite similarly to the probabilistic quantities
for particles. Hence,
in a certain sense, for time analysis along the motion direction,
the wave packet (\ref{eq_phot5}) is quite similar to a wave packet for
nonrelativistic particles and
similarly to the conventional nonrelativistic quantum mechanics, one
can define the same form of time operator as for particles in
nonrelativistic quantum mechanics and hence the mean time and the
distribution variance of times of photon (electromagnetic wave packet)
passing through point $x$ in both time and energy representations)
\cite{Olkhovsky.1997.Trieste}. Then, the same interpretation one can use for the
propagation of electromagnetic wave packets (photons) in media and
waveguides when
reflections and tunneling can take place --- in particular, for
waveguides like depicted in Fig.\ \ref{fig6} with spatially decreasing
and increasing waves in Eq.\ (\ref{eq_phot6}). The only difference is in the
momentum-energy relation (quadratic for particles and linear for photons).

So, from rather simple calculations of $J_{em, x}$ using Eqs.\
(\ref{eq_phot5}) -- (\ref{eq_phot10}), and using the given above definitions
of ${\bf E}$ and ${\bf H}$ (see also \cite{Olkhovsky.1997.JPGCE}), one can
obtain the following relation:
\begin{equation}
   J_{em, x} = Re F(y, z) \biggl(\varphi(x)
        \displaystyle\frac{-iw}{4\pi}
        \displaystyle\frac{\partial\varphi(x)}{\partial x} \biggr),
\label{eq_phot11}
\end{equation}                                  
where the function $F(y, z)$ depends on the boundary conditions of the
waveguide (see Fig.\ \ref{fig6}) and calculated in \cite{Olkhovsky.1997.JPGCE}.
Therefore under boundary conditions the flux density for photons can
be obtained from the flux density for particles by simple replacing
($-i\hbar/2m$) by $F(y, z) [-iw/4\pi]$. At this substitution
all results and relevant expressions (approach to the solution of a
problem on the basis of consideration of multiple internal reflections
of fluxes in the region of the barrier, phase tunneling and reflection
times and other results), obtained above for the description of tunneling
evolution of the particle through the barrier, also take place at the
description of photon propagation.


In the particular case of quasimonochromatic wave packets,
under the same boundary conditions as considered
for the problem of tunneling of a particle through 1D rectangular barrier,
we obtain the identical expression for the phase tunneling time:
\begin{equation}
\begin{array}{ll}
   \tau_{tun, em}^{Ph} = \displaystyle\frac{2}{c\xi_{em}} &
   \mbox{for  } \xi_{em} L >> 1.
\end{array}
\label{eq_phot12}
\end{equation}                                  

From Eq.\ (\ref{eq_phot12}) one can see that when
$\xi_{em} L > 2$ the effective tunneling velocity
\begin{equation}
\begin{array}{ll}
   v_{tun}^{eff} = \displaystyle\frac{L}{\tau_{tun, em}^{Ph}}
\end{array}
\label{eq_phot13}
\end{equation}                                  
is more than $c$, i.e. superluminal. This result agrees with the results
of the microwave-tunneling measurements presented in
\cite{Enders.1992.JPGCE}.

Note, that for sub-barrier energies the nonlocality of a barrier
as a whole takes place not only for nonrelativistic particles but also
for photons. This property is the physical cause of the superluminality
during the tunneling.

\section{Conclusions
\label{sec:level7}}

In this work the method of multiple internal reflections describing
the process of tunneling of a nonrelativistic particles and photons
through barriers of the various forms is presented.
This method is the further development of a series of articles
\cite{Olkhovsky.1992.PRPLC,Olkhovsky.1997.Trieste,Olkhovsky.1997.JPGCE}, devoted to the time description
of tunneling through a barrier. It uses the essentially non-stationary
approach constructed on the basis of multiple reflections (and
transmissions) of w.p. in relation to the boundaries of barrier.
In result one can describe in dependence on time the process of tunneling
of total w.p. describing the considered nonrelativistic particle or
photon, through barrier and to study specific features of process
in any interesting moment of time or in any point of space in details.

The possibility of time description of tunneling through a barrier
is one of the principle perspectives of this method in
comparison with stationary approaches.

The stationary one-dimensional problem of tunneling (and propagation)
of a nonrelativistic particle through a rectangular barrier with
accounting of the multiple internal reflections was earlier solved
\cite{Fermor.1966.AJPIA,McVoy.1967.RMPHA,Anderson.1989.AJPIA}.
For sub-barrier energies the plane waves in the
barrier region (on the basis of which the complete expressions for w.f.
were found) had zero fluxes. According to the physical understanding there
is a problem of applicability of such approach to the problem solution.
In the given article the substantiation of this approach is given on the
basis of using the non-stationary w.p..
For this problem (being the test one) the phase time of tunneling and
reflection in relation to the barrier at whole under solving the
problem on the basis of the method of multiple internal reflections
are introduced.

Using the method of multiple internal reflections the problem
of tunneling of a nonrelativistic particle through a spherically
symmetric barrier is solved for the first time. Here, using this method it
is possible (as against the known stationary approaches) to separate the
wave packet, transmitted through the barrier and describing a particle
after its leaving outside after double tunneling through barrier, from
the wave packet, reflected from the barrier describing a reflected particle
(both packets are spherically divergent). For the diagonal element
of scattering matrix with orbital moment $l$ the following property
\[
S^{l} = S_{tr}^{l} + S_{ref}^{l},
\]
is fulfilled, i.e. the $S$-matrix consists of two components
corresponding to the transmitted and reflected wave packets
in relation to barrier. This property has physical sense and
is proved mathematically.

We suppose, that the method will allow to describe such properties of
nuclear processes, which are not explained by stationary methods.
So, some experiments performed recently, have caused an increased
interest to a bremsstrahlung in an $\alpha$-decay of heavy nuclei
\cite{Kasagi.1997.PRLTA}.
This phenomenon is interesting in that includes both a radiation of
photons in a propagation of an $\alpha$-partiále in an electromagnetic
field of an daughter nucleus, and a tunneling of the $\alpha$-partiále
through the decay barrie.
Now the effect of the photon radiation in the tunneling of the
$\alpha$-partiále under the barrier is investigated unsatisfactoryli.
For a description of this process some stationary methods allowing
to calculate a spectrum of the bremsstrahlung are created.
But in comparison with the experimental data one can see that each
stationary approach describes the phenomenon with a small degree of
an exactitude.
Besides the minima and maximas are registered in the spectrum for some
nuclei, while the stationary methods give a monotonically decreasing
curve for the spectrum.
We assume, that based on a space-time approach the method of multiple
internal reflections will allow to explain the peaks in this spectrum.
A preliminary analysis shows, that these peaks correspond to resonance
levels of the $\alpha$-decay of the researched nucleus and they can be
evaluated using of the method.

In article the possibility to apply the method for 1D problem of photon
tunneling through a rectangular barrier is explored. On the basis of
the given analysis the analogy (having a mathematical substantiation
and physical sense) between wave packets (and also between problem
setting, boundary conditions) describing both propagation and
tunneling of a nonrelativistic particle and photon, is shown. In result,
it is possible to apply the method of multiple internal reflections for
the problem with photons for the first time.
At the found transformation the obtained results for the problem of
particle tunneling through a barrier transform into the relevant
expressions for the problem of tunneling of photons. The tunneling 
durations are found. For enough wide (and high) barrier there is an 
effect of propagation of wave packet with velocity more, than velocity 
of light (Hartman's and Fletcher's effect).

The superluminal phenomena, observed in the experiments presented
in \cite{Enders.1992.JPGCE} and later in other papers (for example,
see the relevant references in
\cite{Olkhovsky.1997.Trieste,Olkhovsky.1997.JPGCE,Jakiel.1998.PHLTA}),
generated a lot of discussions on relativistic causality. And in
connection with this, also an interest for similar phenomena, observed
for the electromagnetic pulse propagation in a dispersive medium
\cite{Chu.1982.PRLTA}, was revived. The known way of usual
understanding consists in explaining the superluminal phenomena during
tunneling on the base
of a pulse attenuated reshaping (or reconstructing) discussed at the
classical limit earlier by
\cite{Chu.1982.PRLTA,Sommerfeld.1907.ZEPYA,Brillouin.1960}:
the later parts of an input pulse are preferentially attenuated in
such a way that the output peak appears shifted toward earlier times,
arising from the forward tail of the incident pulse in a strictly
causal manner \cite{Steinberg.1993.PRLTA}.

Already for long there was ascertained that the wavefront velocity of
the electromagnetic pulse propagation, when pulses have a step-function
envelope, cannot exceed the velocity of light $c$ in vacuum
\cite{Sommerfeld.1907.ZEPYA,Brillouin.1960}. Namely in this the principal demand of the
relativistic (Einstein) causality consists . This conclusion was
confirmed by various methods and in various processes, including
tunneling
\cite{Ranfagni.1995.PHRVA,Mugnai.1995.PHLTA,Fox.1970.PPSOA,Deutch.1993.APNYA,
Hass.1994.PHLTA,Ya.1994.SSCOA,Heitmann.1994.PHLTA}. Note, that it
is known from the momentum-energy Fourier-analysis of an electromagnetic
wave packet with the step-function form of the forward edge, that such
a wave packet contains components with large (up to the infinite) energies,
i.~e. above-barrier energies, for which the superluminality is absent.

One of the argued now problems consists in the absence of a step-function
form of forward edges for realistic wave packets
\cite{Ranfagni.1995.PHRVA,Mugnai.1995.PHLTA,Heitmann.1994.PHLTA}.
In such cases the conclusions of \cite{Sommerfeld.1907.ZEPYA,Brillouin.1960}
can seem to be inapplicable. An interesting approach to analyse the form
of causality namely in such cases was proposed in \cite{Garrison.1998.PHLTA}.

Finally, from the analysis of first step in solving the problem by 
method of multiple internal reflections
one can see that the tunneling process at sub-barrier energies is
a non-local phenomenon becauce during tunneling the entering w.p.
fills up the whole barrier atonce and w.p. feels immediately both
barrier walls (boundaries).

\bibliographystyle{h-physrev4}
\bibliography{JPS_arx}

\newpage
\listoffigures

\newpage

\begin{figure}[p]
\centerline{\includegraphics[width=7cm]{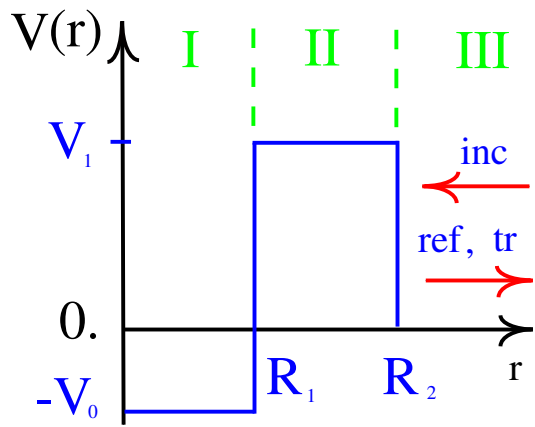}}
\caption{Spherically symmetric rectangular barrier}
\label{fig3}
\end{figure}




\end{sloppypar}
\end{document}